\documentclass[10pt]{iopart}

\usepackage[dvips]{graphicx}
\usepackage{bm}

\newcommand{\w}[1]{\bm{#1}}
 
\begin{document}
 
\title{Rotating star initial data for a 
constrained scheme in numerical relativity}

\author{Lap-Ming Lin and J\'{e}r\^{o}me Novak}
\address{
Laboratoire de l'Univers et de ses Th\'{e}ories, 
Observatoire de Paris, F-92195 Meudon Cedex, France}

\date{\today}
 
\begin{abstract} 
  A new numerical code for computing stationary axisymmetric rapidly rotating
  stars in general relativity is presented. The formulation is based on a
  fully constrained-evolution scheme for $3+1$ numerical relativity using the
  Dirac gauge and maximal slicing. We use both the polytropic and MIT bag
  model equations of state to demonstrate that the code can construct rapidly
  rotating neutron star and strange star models.  We compare numerical models
  obtained by our code and a well-established code, which uses a different
  gauge condition, and show that the two codes agree to high accuracy.

\end{abstract}



\section{\bf Introduction} 
\label{sec:intro}

In the $3+1$ formalism of general relativity, the Einstein equations are
decomposed into a set of four constraint equations and six evolution equations
\cite{york79,bau03}. Solving the (elliptic-type) constraint equations at each
time-step in multidimensional simulations is in general not feasible as it is
computationally expensive for most numerical techniques.  Hence, a
free-evolution approach, i.e., solving the constraint equations only for the
initial data and performing the evolutions without enforcing the constraints,
is generally favoured over a fully constrained scheme (solving the constraints
at each time step) in three-dimensional numerical simulations.  While
mathematically the constraints are preserved by the evolution equations, in
practice small constraint violations due to numerical errors typically grow
quickly to a significant level that make the solution unphysical and plague
the simulations.  Although numerous techniques to control the growth of
constraint violations have been developed (e.g.,
\cite{cal02,lin04,kid05,hol04,tig04,gen04}), it is not clear yet to what
extent they can control the constraint violations successfully in general.

Fully constrained-evolution scheme has been used in the past only in
spherically symmetric or axisymmetric problems (e.g.,
\cite{sta85,cho93,abr94,cho03}).  The main advantage of a fully constrained
scheme is that the constraints are fulfilled to within the discretisation
errors, and the constraint-violating modes do not exist by construction.
Recently, a new formulation for $3+1$ numerical relativity based on a fully
constrained-evolution scheme is proposed by Bonazzola {\it et al.}
\cite{bon04}.  The chosen coordinate conditions (maximal slicing and Dirac
gauge, see Sec.~\ref{sec:metric}), together with the use of spherical
components of tensor fields, reduce the ten Einstein equations to a system of
five quasi-linear elliptic equations, which are solved by efficient multi-domain
spectral methods, and two quasi-linear scalar wave equations \cite{bon04}. The
Dirac gauge is used to fix the remaining three degrees of freedom.  The
stability of the proposed scheme has been demonstrated for a three-dimensional
pure gravitational wave spacetime \cite{bon04}.

In the proposed constrained scheme, the coordinates are fixed by the Dirac
gauge on each hypersurface (with maximal slicing condition), including the
initial one.  This implies that initial data must be prepared in the same
coordinate choices in order to perform dynamical evolutions. As an advantage
of this gauge-fixing, stationary solutions of the Einstein equations can be
computed within the same framework simply by setting the time derivative terms
to zero in the equations.  The aim of the present work is to construct
rotating-star initial data for this new formulation of $3+1$ numerical
relativity. For this purpose, we have developed a numerical code to calculate
stationary axisymmetric rotating star models based on the Dirac gauge and
maximal slicing.

The purpose of this paper is to present the formulation of the problem and the
tests that have been done to validate our numerical code. The numerical code
can be used to provide initial rotating star models for hydrodynamics
simulations in full general relativity within the new formulation.  Our
emphasis here is put on the comparison of the accuracy between our code and a
well-established code LORENE-rotstar \cite{bon93,gou99,lorene}. In particular,
we demonstrate that our numerical code can compute rapidly rotating neutron
star and strange star models to high accuracy.  Unless otherwise noted, we use
units such that $G=c=1$. Latin (Greek) indices go from 1 to 3 (0 to 3).

\section{Formulation}

\subsection{The $3+1$ decomposition}
\label{sec:3+1}

In this section, we give a brief description of the $3+1$ formulation of the
Einstein equations in order to define our notations (see, e.g.,
\cite{york79,bau03} for details).  In the $3+1$ formalism, the spacetime is
foliated by a family of spacelike hypersurfaces $\Sigma_t$, labelled by the
time coordinate $t$. Introducing a coordinate system $(x^i)$ on each
hypersurface, the line element may be written as
\begin{equation}
ds^2 = -N^2 dt^2 + \gamma_{ij} (dx^i + \beta^i dt)
        (dx^j + \beta^j dt) ,
\end{equation}
where $N$ is the lapse function, $\beta^i$ is the shift vector, and
$\gamma_{ij}$ is the 3-metric induced by the spacetime metric
$g_{\alpha\beta}$ onto each hypersurface $\Sigma_t$
\begin{equation}
\gamma_{\alpha\beta} := g_{\alpha\beta} + n_{\alpha} n_{\beta} .
\end{equation}
Here $n_{\alpha} := -N \nabla_{\alpha} t$ is the unit normal to $\Sigma_t$, 
where $\nabla_{\alpha}$ is the covariant derivative associated with the 
spacetime metric $g_{\alpha\beta}$. 
The stress-energy tensor $T^{\alpha\beta}$ is decomposed as 
\begin{equation}
T^{\alpha\beta} = E n^{\alpha}n^{\beta} + n^{\alpha} J^{\beta} + 
J^{\alpha} n^{\beta} + S^{\alpha\beta} , 
\label{eq:Tmunu_decomp}
\end{equation}
where $E := T_{\alpha\beta}n^{\alpha}n^{\beta}$, $J_{\alpha} := -
\gamma_{\alpha}^{\ \mu}T_{\mu\nu}n^{\nu}$, and $S_{\alpha\beta} :=
\gamma_{\alpha}^{\ \mu} \gamma_{\beta}^{\ \nu}T_{\mu\nu}$ are the energy 
density, momentum density, and the stress tensor, as measured by the 
observers of 4-velocity $n^{\alpha}$ (the so-called Eulerian observers $O_e$).

The evolution of the 3-metric
$\gamma_{ij}$ is governed by 
\begin{equation}
{\partial\over \partial t}\gamma_{ij} - {\cal L}_{\w{\beta}} \gamma_{ij}
= - 2N K_{ij} , 
\label{eq:adm_dgamma_dt}
\end{equation}
where $\cal L$ is the Lie derivative operator and $K_{ij}$ is the 
extrinsic curvature of $\Sigma_t$. The evolution equation for $K_{ij}$ 
is
\begin{eqnarray}
{\partial\over \partial t} K_{ij} - {\cal L}_{\w{\beta}} K_{ij}
&=& - D_i D_j N + N \{ R_{ij} - 2 K_{ij} K^{ij} + K K_{ij} \cr
&& \cr
&& 
 + 4\pi [ (S-E)\gamma_{ij} - 2S_{ij} ] \} ,
\end{eqnarray}
where $D_i$ is the covariant derivative associated with the 3-metric 
$\gamma_{ij}$, $R_{ij}$ is the Ricci tensor associated with this 3-metric, 
$K := K_{\ i}^i$ is the trace of the extrinsic curvature, and 
$S := S_{\ i}^i$. 
In the 3+1 formulation, the full set of Einstein equations is equivalent to  
the above evolution equations, together with the Hamiltonian constraint 
\begin{equation}
R + K^2 - K_{ij}K^{ij} = 16\pi E ,
\label{eq:ham_const}
\end{equation}
and the momentum constraint 
\begin{equation}
D_j K_i^{\ j} - D_i K = 8\pi J_i ,
\label{eq:mom_const} 
\end{equation}
where $R :=R^i_{\ i}$ is the three-dimensional Ricci scalar.

\subsection{The matter sources: uniformly rotating fluid}
\label{sec:matter}

We assume that the matter consists of a perfect fluid with a stress-energy
tensor 
\begin{equation}
T^{\alpha\beta} = (e+P)u^{\alpha}u^{\beta} + P g^{\alpha\beta} ,
\end{equation}
where $u^{\alpha}$ is the 4-velocity of the fluid; $e$ and $P$ are 
respectively the energy density and pressure, as measured by the 
the fluid comoving observer $O_f$. Defining the Lorentz factor 
linking the two observers $O_e$ and $O_f$ by 
\begin{equation}
\Gamma := -n_{\alpha} u^{\alpha} = N u^{t} , 
\label{eq:Gamma}
\end{equation}
the energy density $E$ in Eq.~(\ref{eq:Tmunu_decomp}) can be written as
\begin{equation}
E = \Gamma^2 (e+P) - P .
\end{equation} 
The momentum density $J^i$ is 
\begin{equation}
J^{i} = (E+P) v^i , 
\end{equation} 
where the fluid 3-velocity $v^i$ is related to the spatial components of the 
fluid 4-velocity $u^i$ by 
\begin{equation}
u^i = \Gamma \left ( v^i - {\beta^i\over N} \right) .
\label{eq:4_velocity}
\end{equation}
Note that the Lorentz factor can be expressed as 
$\Gamma = (1- v^2)^{-1/2}$, where $v := (v_i v^i)^{1/2}$ is 
the ``physical'' fluid speed as measured by the Eulerian observer $O_e$. 
The stress tensor $S_{ij}$ is given by 
\begin{equation}
S_{ij} = (E+P) v_i v_j + P \gamma_{ij} . 
\end{equation}

Up to this point, we have not made any assumptions on the spacetime and fluid
flow. Now we consider that the spacetime is stationary, axisymmetric, and
asymptotically flat. These assumptions imply the existence of two Killing
vector fields: $\vec{\zeta}$, which is timelike at spatial infinity;
$\vec{\chi}$, which is spacelike everywhere and with closed orbits.
Furthermore, the Killing vectors commute \cite{car70}, and hence we can choose
an adapted coordinate system $(t,x^1,x^2,\varphi)$, such that $\vec{\zeta} =
\partial / \partial t$ and $\vec{\chi} = \partial / \partial \varphi$ are the
coordinate vector fields (see also \cite{bon93}). We choose the remaining two
coordinates to be of spherical type (i.e., $x^1=r$ and $x^2=\theta$).  

We also impose the so-called circularity condition on the stress-energy tensor:
\begin{eqnarray}
T^\alpha_{\ \beta}\zeta^\beta &=& \mu \zeta^\alpha + \nu \chi^\alpha , \\
&& \cr
T^\alpha_{\ \beta}\chi^\beta &=& \lambda \zeta^\alpha + \sigma \chi^\alpha . 
\end{eqnarray}
This condition is equivalent to the absence of meridional convective
currents, and implies that the fluid 4-velocity is given by
\begin{equation}
\vec{u} = u^t \left( {\partial\over \partial t} 
+ \Omega {\partial\over \partial \varphi} \right) , 
\end{equation}
where $\Omega := u^{\varphi}/u^t$ is the fluid coordinate angular 
velocity, and can be interpreted as the fluid angular velocity 
as seen by an inertial observer at rest at infinity. A theorem of Carter  
\cite{car69} shows that the circularity condition implies that the line 
element can be written as 
\begin{equation}
ds^2 = -N^2 dt^2 + \gamma_{\varphi\varphi}(d\varphi + \beta^\varphi dt)^2
+ \gamma_{rr} dr^2 + 2 \gamma_{r\theta} dr d\theta + \gamma_{\theta\theta} d\theta^2 .
\end{equation}
Notice that only the $\varphi$-component of the shift vector is nonzero and 
we have not specified the gauge choice at this point. The so-called 
quasi-isotropic gauge (see Sec.~\ref{sec:tests}) corresponds to 
$\gamma_{r\theta}=0$, $\gamma_{\theta\theta} = r^2\gamma_{rr}$, while 
the Dirac gauge relates the metric components by differential 
equations (see Eq.~(\ref{eq:dirach})).

The equation of stationary motion follows from the projection of the
conservation equation $\nabla_{\alpha} T^{\alpha\beta}=0$ normal to the
4-velocity $u^{\alpha}$.  In this paper, we focus on the case of uniformly
rotating star (i.e., $\Omega$ is a constant). We also assume that the fluid 
is barotropic. 
In this case, the equation of stationary motion can be integrated analytically 
and is given by (see, e.g., \cite{bon93})
\begin{equation}
H + \ln N - \ln \Gamma = {\rm const.} ,
\label{eq:fluid_1st}
\end{equation}
where $H$ is the log-enthalpy defined by 
\begin{equation}
H := \int { dP\over e + P} . 
\label{eq:H}
\end{equation}

\subsection{The metric equations}
\label{sec:metric}

Here we summarise the full set of Einstein equations in the
constrained-evolution scheme based on the Dirac gauge and maximal slicing.  We
refer the reader to Ref.~\cite{bon04} for details and derivations.  First, we
define a conformal metric $\tilde{\gamma}_{ij}$ by
\begin{equation}
\tilde{\gamma}_{ij} := \Psi^{-4} \gamma_{ij} , 
\label{eq:conf_gam}
\end{equation}
with the conformal factor $\Psi$ defined by 
\begin{equation}
\Psi := \left( {\det \gamma_{ij}\over \det f_{ij} } \right)^{1/12} ,
\label{eq:conf_psi}
\end{equation}
where $f_{ij}$ is a flat 3-metric, given by the asymptotic condition on
$\gamma_{ij}$.  The four constraint equations (\ref{eq:ham_const}) and
(\ref{eq:mom_const}), together with the maximal slicing condition $K=0$,
result in two scalar equations for the lapse and conformal factor; and one
vectorial elliptic equations for the shift vector.

The lapse function $N$ is given by 
\begin{equation}
\Delta N = \Psi^4 N \left[ 4\pi (E+S) + \tilde{A}_{kl} A^{kl} \right]
 - h^{kl}{\cal D}_k{\cal D}_l N - 2 \tilde{D}_k \Phi \tilde{D}^k N ,
\label{eq:N_eq}
\end{equation}
where ${\cal D}_i$ is the covariant derivative associated with the flat metric
$f_{ij}$ and its contravariant component is defined by ${\cal D}^i := f^{ij}
{\cal D}_j$; $\Delta := f^{ij} {\cal D}_i {\cal D}_j$ is the flat-space
Laplacian operator; $\tilde{D}_i$ is the covariant derivative associated with
the conformal metric $\tilde{\gamma}_{ij}$ and its contravariant component is
$\tilde{D}^i := \tilde{\gamma}^{ij} \tilde{D}_j$ (with the inverse conformal
metric $\tilde{\gamma}^{ij}$ defined by $\tilde{\gamma}_{ik}
\tilde{\gamma}^{kj} = \delta_i^{\ j}$). The quantity $\Phi$ is defined by
$\Phi := \ln \Psi$.  The traceless part of the conformal extrinsic curvature
$A^{ij}$ is defined by
\begin{equation}
A^{ij} := \Psi^4 \left( K^{ij} - {1\over 3} \gamma^{ij} K \right) ,
\label{eq:Aij}
\end{equation}
while the tensor field $\tilde{A}_{ij}$ is defined by 
\begin{equation}
\tilde{A}_{ij} := \tilde{\gamma}_{ik} \tilde{\gamma}_{jl} A^{kl} 
= \Psi^{-4} \left( K_{ij} - {1\over 3} \gamma_{ij} K \right) . 
\end{equation}
The tensor field $h^{ij}$ on the right-hand side (RHS) of Eq.~(\ref{eq:N_eq})
is the deviation of the inverse conformal metric from the inverse flat metric
defined by
\begin{equation}
h^{ij} := \tilde{\gamma}^{ij} - f^{ij} . 
\end{equation}
In the proposed constrained scheme, the Dirac gauge condition is given by 
\begin{equation}
{\cal D}_j h^{ij} = 0 . 
\label{eq:dirach}
\end{equation}

Next, the conformal factor $\Psi$ (or equivalently $Q := \Psi^2 N$) 
is determined from 
\begin{eqnarray}
\Delta Q &=& - h^{kl}{\cal D}_k{\cal D}_l Q + \Psi^6\left[
N\left( 4\pi S + {3\over 4}\tilde{A}_{kl}A^{kl} \right)\right] \cr
&& \cr
&& 
+ 2\Psi^2\left[ N\left( {\tilde{R}_*\over 8} + \tilde{D}_k\Phi
\tilde{D}^k\Phi \right) + \tilde{D}_k\Phi \tilde{D}^k N \right] ,
\label{eq:Q_eq}
\end{eqnarray}
where the quantity $\tilde{R}_*$ on the RHS is given by 
\begin{equation}
\tilde{R}_* := {1\over 4} \tilde{\gamma}^{kl}{\cal D}_k h^{mn}
{\cal D}_l\tilde{\gamma}_{mn} - {1\over 2}\tilde{\gamma}^{kl}
{\cal D}_k h^{mn} {\cal D}_n\tilde{\gamma}_{ml} .
\end{equation}
The elliptic equation for the shift vector is 
\begin{eqnarray}
\fl \Delta \beta^i + {1\over 3}{\cal D}^i({\cal D}_j \beta^j)
= 16\pi N\Psi^4 J^i + 2A^{ij}{\cal D}_j N 
-12N A^{ij}{\cal D}_j\Phi - 2N \Delta^i_{\ kl} A^{kl} \nonumber \\ 
-h^{kl}{\cal D}_k{\cal D}_l \beta^i - {1\over 3}h^{ik}{\cal D}_k
{\cal D}_l \beta^l ,
\label{eq:beta_eq}
\end{eqnarray}
where the tensor field $\Delta^k_{\ ij}$ is defined by 
\begin{equation}
\Delta^k_{\ ij} := {1\over 2}\tilde{\gamma}^{kl}\left ( 
{\cal D}_i \tilde{\gamma}_{lj} + {\cal D}_j \tilde{\gamma}_{il}
- {\cal D}_l \tilde{\gamma}_{ij} \right ) .
\label{eq:Delta_ijk}
\end{equation}

Now we turn to the dynamical part of the Einstein equations. In the proposed
constrained scheme, one solves for the tensor field $h^{ij}$ instead of
$\tilde{\gamma}^{ij}$. The evolution equation in this formulation is given by
a flat-space (second-order) wave equation for $h^{ij}$ (see Eq. (85) of
Ref.~\cite{bon04}). As mentioned in Sec.~\ref{sec:intro}, one advantage of
using the Dirac gauge to fix the coordinates on each slice $\Sigma_t$ is that
stationary solutions of the Einstein equations can be computed within the same
scheme, simply by setting the time-derivative terms to zeros in the equations.
This is possible because of the existence of the Killing vector
$\vec{\zeta}=\partial/\partial t$ as discussed in Sec.~\ref{sec:matter}. This
reduces the wave equation for $h^{ij}$ to the following tensorial Poisson-like
equation:
\begin{eqnarray}
\fl \Delta h^{ij} = {\Psi^4\over N^2} \left\{ 
{\cal L}_{\w{\beta}} {\cal L}_{\w{\beta}} h^{ij} +
{4\over 3}{\cal D}_k\beta^k  {\cal L}_{\w{\beta}} h^{ij} +
{N\over \Psi^6}{\cal D}_k Q \left( {\cal D}^i h^{jk} + {\cal D}^j h^{ik}
- {\cal D}^k h^{ij} \right) - 2N {\cal S}^{ij} \right. \nonumber \\
 \left. 
- {2\over 3}{\cal D}_k\beta^k (L\beta)^{ij}  
+ 2 ( {\cal L}_{\w{\beta}} N ) A^{ij}  
+ {2\over 3} \left[ {\cal L}_{\w{\beta}} ( {\cal D}_k \beta^k )
+ {2\over 3} ({\cal D}_k\beta^k)^2 \right] h^{ij} \right. \nonumber \\
\left.
- {\cal L}_{\w{\beta}} (L\beta)^{ij} \right\} ,
\label{eq:hij}
\end{eqnarray}
where the notation $(L \beta)^{ij}$ stands for the conformal Killing operator 
associated with the flat metric acting on the shift vector $\beta^i$:
\begin{equation}
(L\beta)^{ij} := {\cal D}^i\beta^j + {\cal D}^j\beta^i 
-{2 \over 3} {\cal D}_k\beta^k f^{ij} .
\end{equation}
The tensor field ${\cal S}^{ij}$ on the RHS of Eq. (\ref{eq:hij}) 
is given by 
\begin{eqnarray}
\fl {\cal S}^{ij} = \Psi^{-4}\left \{ N\left( \tilde{R}_*^{ij} + 
8\tilde{D}^i\Phi \tilde{D}^j\Phi \right) + 4\left( \tilde{D}^i\Phi
\tilde{D}^j N + \tilde{D}^j\Phi \tilde{D}^i N \right) \right. \nonumber \\
\left. 
- {1\over 3}\left[ N\left( \left[ \tilde{R}_* + 8\tilde{D}_k\Phi\tilde{D}^k\Phi
\right] \tilde{\gamma}^{ij}\right) + 8\tilde{D}_k\Phi \tilde{D}^k N 
\tilde{\gamma}^{ij} \right] \right \} \nonumber \\
+ 2N \left[ \tilde{\gamma}_{kl}A^{ik}A^{jl} - 4\pi\left(\Psi^4 S^{ij}
-{1\over 3}S \tilde{\gamma}^{ij} \right) \right] \nonumber \\
- \Psi^{-6}\left[ \tilde{\gamma}^{ik}\tilde{\gamma}^{jl} {\cal D}_k
{\cal D}_l Q 
+ {1\over 2} \left( h^{ik}{\cal D}_k h^{lj} + h^{kj}{\cal D}_k h^{il}
- h^{kl}{\cal D}_k h^{ij} \right) {\cal D}_l Q  \right. \nonumber \\
\left.
- {1\over 3}\left(\tilde{\gamma}^{kl}{\cal D}_k{\cal D}_l Q 
\tilde{\gamma}^{ij} \right) \right] , 
\label{eq:cal_Sij}
\end{eqnarray}
with 
\begin{eqnarray}
\fl \tilde{R}_*^{ij} = {1\over 2}\left[ h^{kl} {\cal D}_k{\cal D}_l h^{ij}
- {\cal D}_l h^{ik} {\cal D}_k h^{jl} - \tilde{\gamma}_{kl}
\tilde{\gamma}^{mn}{\cal D}_m h^{ik} {\cal D}_n h^{jl}  \right. \nonumber \\
\left. 
+\tilde{\gamma}_{nl}{\cal D}_k h^{mn} \left(\tilde{\gamma}^{ik}
{\cal D}_m h^{jl} + \tilde{\gamma}^{jk} {\cal D}_m h^{il} \right)
+ {1\over 2} \tilde{\gamma}^{ik} \tilde{\gamma}^{jl}{\cal D}_k h^{mn}
{\cal D}_l \tilde{\gamma}_{mn} \right] .
\label{eq:tRij*}
\end{eqnarray}
Furthermore, after setting the time-derivative term of $h^{ij}$ to zero, 
the conformal extrinsic curvature $A^{ij}$ as defined in 
Eq.~(\ref{eq:Aij}) is deduced from (see Eq. (92) of \cite{bon04})
\begin{equation}
A^{ij} = {1\over 2N} \left[ (L\beta)^{ij} 
-{\cal L}_{\w{\beta}} h^{ij} - {2\over 3}{\cal D}_k\beta^k h^{ij} \right] .
\label{eq:Aij_solve}
\end{equation}

In summary, to calculate stationary axisymmetric uniformly rotating star
models in the framework of the constrained scheme \cite{bon04} based on the
Dirac gauge and maximal slicing, one needs to solve for two scalar elliptic
equations (\ref{eq:N_eq}) and (\ref{eq:Q_eq}) respectively for $N$ and $\Psi$,
a vectorial Poisson-like equation (\ref{eq:beta_eq}) for $\beta^i$, and a
tensorial elliptic equation (\ref{eq:hij}) for $h^{ij}$, together with the
first integral of motion Eq.~(\ref{eq:fluid_1st}) for the matter. The
numerical procedure on how to solve this system of equations is described in
Sec.~\ref{sec:procedure}.


\subsection{Global quantities}
\label{sec:global}

We list here various global quantities relevant to axisymmetric rotating-star
spacetimes.  These gauge invariant quantities are useful to estimate the
accuracy of our numerical code as they provide a direct comparison between our
code and a different code, which uses a different gauge condition, as
presented in Sec.~\ref{sec:tests}.

Given a baryon current $n u^{\alpha}$, where $n$ is the number density 
in the fluid frame, the baryon mass of the star is expressed as  
\begin{equation}
M_b = m_{\rm B} \int \left[ -n_{\alpha} \left( n u^{\alpha} \right)\right]
  dV = m_{\rm B} \int \Gamma n dV , 
\end{equation}
where $m_{\rm B}$ is the baryon rest mass, $dV = \sqrt{\gamma} d^3x$ is the
proper 3-volume element (with $\gamma$ being the determinant of the 3-metric),
and we have used Eq.~(\ref{eq:Gamma}) in the second equality. As we follow
Ref. \cite{bon04} to expand all tensor fields onto the spherical basis $(
\w{e}_{\hat i}) = ( {\partial\over\partial r}, {1\over
  r}{\partial\over\partial \theta}, {1\over r\sin\theta}{\partial\over\partial
  \varphi} )$, which is orthonormal with respect to the flat metric, the
proper volume element is written explicitly as $dV = \sqrt{\hat{\gamma}}
r^2\sin\theta dr d\theta d\varphi$.  Notice that we denote by $\hat\gamma$ the
determinant of the 3-metric expanded onto the basis $( \w{e}_{\hat i})$.  Here
and afterward we denote by $(\hat r, \hat \theta, \hat \varphi)$ indices of
specific components on the orthonormal basis $(\w{e}_{\hat i})$.

The gravitational mass $M_g$ is given by the Komar integral 
(see Eq. (11.2.10) of \cite{wald84}) 
\begin{equation}
M_g = 2 \int \left( T_{\alpha\beta} - {1\over 2} T_{\ \lambda}^\lambda
g_{\alpha\beta} \right) n^{\alpha} \zeta^{\beta} dV ,
\end{equation}
where $\zeta^{\alpha}$ is the timelike 
Killing vector discussed in Sec.~\ref{sec:matter}. Explicitly, we have 
\begin{equation}
M_g = \int \left[ N (E + S) - 2 J_{\hat \varphi} \beta^{\hat\varphi}
\right] dV , 
\end{equation}
The total angular momentum $J$ is given by (see Problem 6 of \cite{wald84})
\begin{equation}
J = - \int T_{\alpha\beta} n^{\alpha} \chi^{\beta} dV
= \int  J_{\hat \varphi} r \sin\theta  dV ,
\end{equation}
where $\chi^{\alpha}$ is the axial Killing vector of the spacetime. 
The rotational kinetic energy for a uniformly rotating star is 
$T = {1\over 2} \Omega J$. The gravitational potential energy is 
\begin{equation}
W = M_p + T - M_g ,
\end{equation}
where $M_p$ is the proper mass of the star defined by 
\begin{equation}
M_p = \int \left[ - n_\alpha \left( e u^\alpha \right) \right] dV
= \int \Gamma e dV . 
\end{equation}
Note that $W$ is defined to be positive.

Furthermore, two relativistic virial identities (the so-called GRV2 and GRV3)
have been proved to be useful for checking the consistency and accuracy of
numerical solutions of rotating relativistic star models.  The
three-dimensional virial identity GRV3 \cite{grv3} is a relativistic
generalisation of the Newtonian virial identity, valid for any stationary and
asymptotically flat spacetime. The two-dimensional virial identity GRV2
\cite{grv2} is valid for any asymptotically flat spacetime (without any
symmetry assumption).  The two virial identities are integral relations
between the matter and metric fields (see \cite{grv3,grv2} for expressions).
In practice, we define the quantities $GRV2:= |1-\lambda_2|$ and $GRV3 :=
|1-\lambda_3|$ as the error indicators for the virial identities, where
$\lambda_2$ and $\lambda_3$ are defined via the integral relations, such that
exact solutions of the Einstein equations satisfy $GRV2=GRV3=0$ (see
\cite{noz98}).  Note that these identities are not imposed during the
numerical calculation, and hence are useful indicators for checking the
accuracy of numerical results.

\section{Numerical results}

\subsection{Numerical procedure}
\label{sec:procedure}

To calculate stationary axisymmetric rotating stars within the Dirac gauge and
maximal slicing, we solve the nonlinear elliptic equations described in
Sec.~\ref{sec:metric} iteratively by means of multi-domain spectral methods
\cite{bon98,bon99} in spherical coordinates.  The code is constructed upon the
LORENE C++ library \cite{lorene}.  We use three spherical numerical domains to
cover the whole hypersurface $\Sigma_t$.  Specifically, we use one domain to
cover the whole star and one domain for the space around the star (typically
to about twice the stellar radius).  The last domain covers the space out to
spatial infinity by means of a compactification $u=1/r$ \cite{bon93}.

In each domain, we use $N_\theta$ collocation points in the polar direction
and $N_\varphi=1$ point in the azimuthal direction for the spectral method.
For the radial direction, we can choose to have different numbers of
collocation points in different domains. In Sec.~\ref{sec:tests}, we use the
notation $N_r=(N_{r1}, N_{r2}, N_{r3})$, where $N_{r1}$ denotes the number of
points in the first domain etc, to specify the grid structure in the radial
direction.

The numerical iteration procedure is briefly described here.  For a given
equation of state, we choose $\Omega$ and the central value of the
log-enthalpy $H$ (see Eq.~(\ref{eq:H})) as the physical parameters that
specify the rotating star model. First we start with an initial guess by
setting all the metric quantities to their flat spacetime values, together
with a spherically symmetric distribution for the matter sources.  The
iteration procedure begins by solving Eqs.~(\ref{eq:N_eq}) and
(\ref{eq:beta_eq}) respectively for the corresponding lapse and shift.  Thus,
we obtain the only nonzero component of the 3-velocity $v^{\hat\varphi} =
(\Omega r \sin\theta + \beta^{\hat\varphi} )/ N$ (see
Eq.~(\ref{eq:4_velocity})), and hence the Lorentz factor $\Gamma$. Next we use
the first integral of motion (Eq.~(\ref{eq:fluid_1st})) to obtain $H$, from
which we deduce the pressure $P$ and the energy density $e$ through the EOS.
Finally, we solve Eqs.~(\ref{eq:Q_eq}) and (\ref{eq:hij}) respectively for $Q$
and $h^{ij}$. The iteration procedure continues until the relative difference
in $H$ throughout the whole star between two consecutive steps is smaller than
some prescribed value.

The resolutions of the scalar Poisson equations for $N$ and $Q$, and the
vectorial elliptic equation for $\beta^i$ have been described in details in
\cite{gra01}. The technique for solving the tensorial Poisson
equation~(\ref{eq:hij}) is described in \ref{sec:appen}.

\subsection{Equation of state}
\label{sec:eos}

In Sec.~\ref{sec:tests}, we present various tests that have been done to
validate our numerical code. For this purpose, we use a polytropic EOS in the
following form to construct rotating neutron star models:
\begin{equation}
P = \kappa n^{\gamma} , 
\end{equation}
where $\kappa$ and $\gamma$ are constants. The number density
$n$ is related to the energy density $e$ by 
\begin{equation} 
e = m_{\rm B} n + {\kappa\over \gamma - 1} n^{\gamma} ,
\end{equation}
where the baryon mass $m_{\rm B} = 1.66\times 10^{-24}\ {\rm g}$. In 
particular, we take $\gamma=2$ and $\kappa=0.03 \rho_{\rm nuc} c^2 /
n_{\rm nuc}^2$, where $\rho_{\rm nuc} = 1.66\times 10^{14}\ {\rm g\ cm^{-3}}$
and $n_{\rm nuc}=0.1\ {\rm fm^{-3}}$. For this EOS, the log-enthalpy is 
given analytically by 
\begin{equation}
H = \ln \left[ 1 + {\kappa \gamma\over m_{\rm B} (\gamma-1) }
n^{\gamma - 1} \right] . 
\end{equation}

We also use the simplest MIT bag model EOS, with noninteracting massless
quarks, to construct rapidly rotating strange stars.  The EOS is given in the
following form (see, e.g., \cite{gou99})
\begin{eqnarray}
P &=& {1\over 3} a n^{4/3} - B , \cr 
&& \cr
e &=& a n^{4/3} + B ,
\end{eqnarray}
where $B$ is the MIT bag constant and the parameter $a = 9 \pi^{2/3} \hbar c /
4 = 952.371\ {\rm MeV\ fm}$.  We choose $B=60\ {\rm MeV\ fm^{-3}}$ in this
work. The stellar surface is characterised by the properties of strange matter
at zero pressure: the number density $n_0 = 0.28665\ {\rm fm^{-3}}$ and the
mass density $\rho_0 = 4.2785\times 10^{14}\ {\rm g\ cm^{-3}}$.  The
log-enthalpy is related to $n$ simply by $H = \ln (n/n_0)^{1/3}$.  The MIT bag
model EOS is useful to test our numerical code in the highly relativistic
regime, since strange stars can reach higher compactness ratios and rotation
rates than ordinary neutron stars.

\begin{figure}
\centering
\includegraphics*[width=8cm]{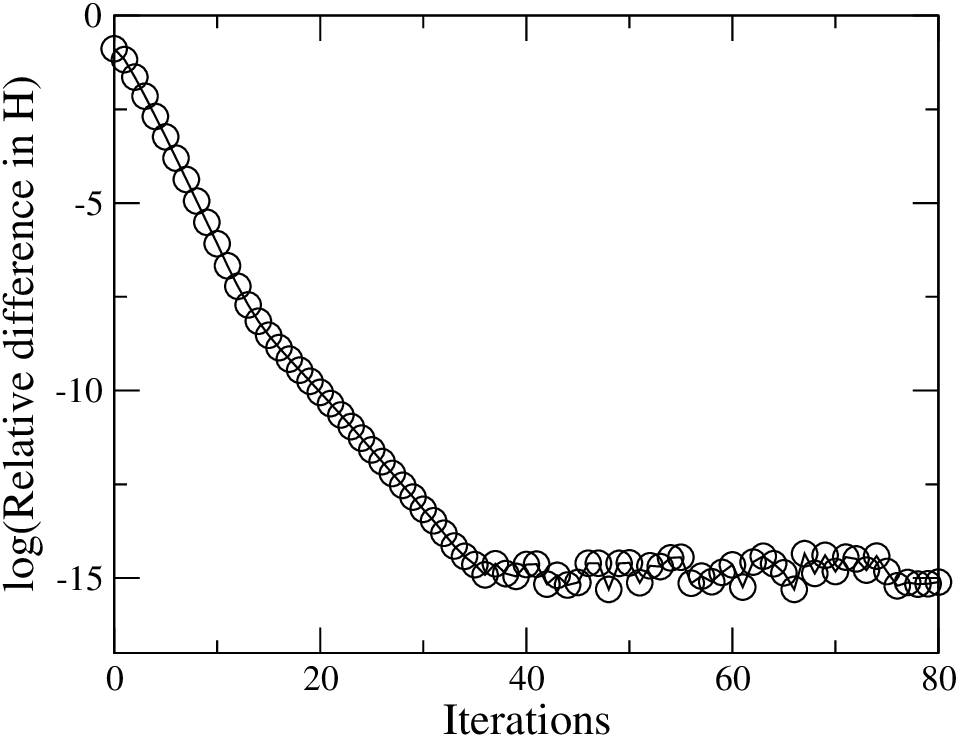}
\caption{Convergence towards zero of the relative difference in $H$ 
throughout the star between two consecutive steps for a non-rotating 
polytropic star model (see text).}
\label{fig:polyNR_diffH}
\end{figure}

\subsection{Tests of the numerical code}
\label{sec:tests}

To test our numerical code, we start with a non-rotating star modelled by the
polytropic EOS described in Sec.~\ref{sec:eos}. The central value of the
log-enthalpy is $H_0=0.2308$ (or equivalently the energy density $e_0=4.889
\rho_{\rm nuc} c^2$) and the baryon mass of the star is $M_b = 1.6 M_{\odot}$.
The star has a gravitational mass $M_g=1.4866 M_{\odot}$ and a compactness
ratio $M_g/R=0.147$, where $R$ is the circumferential radius. In the numerical
calculations, we use a parameter $\epsilon_H$ (typically set to be $10^{-10}$
or smaller) to control the iteration procedure and the precision of the
numerical models: the iteration (see Sec.~\ref{sec:procedure}) is stopped if
the relative difference in $H$ throughout the whole star between two
consecutive steps is smaller than $\epsilon_H$.  In
Fig.~\ref{fig:polyNR_diffH}, we show the convergence of the relative
difference in $H$ towards zero with the number of iterations using radial
collocation points $N_r=(33,33,17)$.  We see that a precision of $10^{-15}$ is
achieved for the numerical result within 40 iteration steps. After that, the
accuracy is limited by the round-off errors. The solution also satisfies the
virial identities to the level of $10^{-15}$.

\begin{figure}
\centering
\includegraphics*[width=8cm]{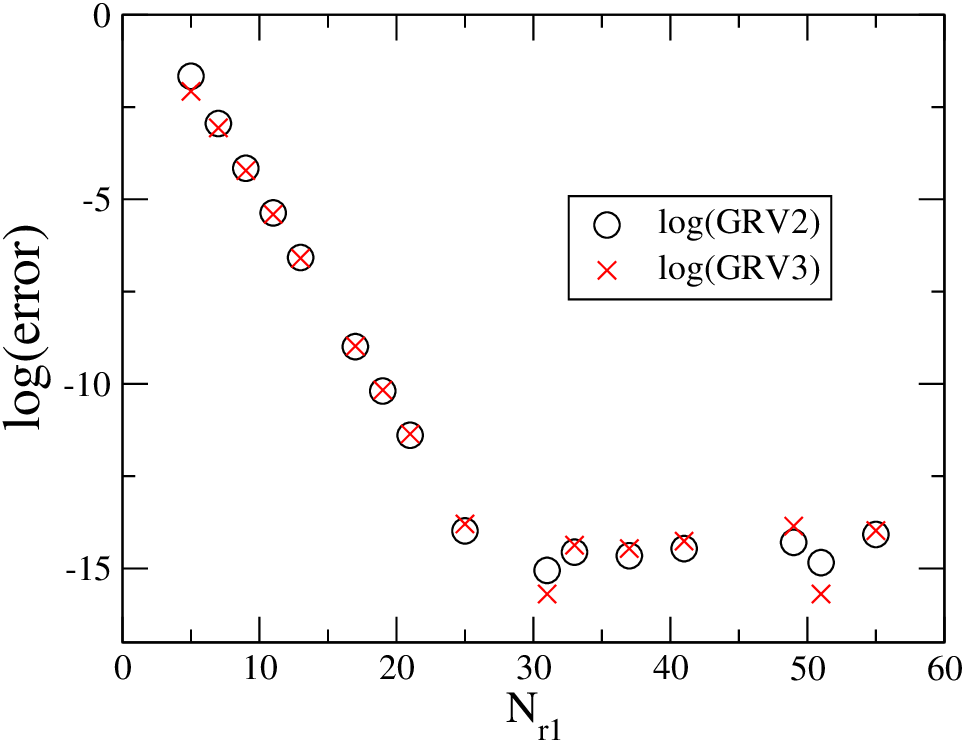}
\caption{Convergence towards zero of $GRV2$ and $GRV3$ with the number 
of collocation points $N_{r1}$ inside the star for the same model as 
shown in Fig.~\ref{fig:polyNR_diffH}.}
\label{fig:polyNR_grv_convg}
\end{figure}

Next, we test the convergence property of the numerical code with respect to
increasing number of radial collocation points using the same model. In
particular, we vary the number of points in the first numerical domain (i.e.,
inside the star), while keeping the points in the other two domains fixed to
$N_{r2}=33$ and $N_{r3}=17$, and we choose $\epsilon_H=10^{-15}$ in this test.
In Fig.~\ref{fig:polyNR_grv_convg}, we plot $\log (GRV2)$ and $\log (GRV3)$
together against the number of points $N_{r1}$ in the first domain.  It is
seen clearly that both $GRV2$ and $GRV3$ converge exponentially towards zero
with the number of points, as expected for spectral methods. The accuracy is
limited by the round-off errors for $N_{r1} > 30$.

Starting from the above non-rotating polytropic model, we then construct a
sequence of increasing uniformly rotating configurations at fixed baryon mass
$M_b=1.6 M_{\odot}$, up to the mass-shedding limit. The accuracy of the
numerical models are estimated by comparing the results to those obtained by a
separated code LORENE-rotstar \cite{bon93,gou99}, which uses the so-called
quasi-isotropic gauge to construct rotating relativistic stars.
LORENE-rotstar is a well-established code which has been tested extensively
and compared with a few different numerical codes \cite{noz98}.  A comparison
of some of the gauge invariant quantities for the sequence between our code
and LORENE-rotstar is given in Table~\ref{tab:poly}. The parameters used to obtain the
numerical results are $N_r=(33,17,17)$, $N_\theta=17$, and
$\epsilon_H=10^{-10}$ for both codes.  In the table, for each value of the
rotation frequency $f$, we display the results obtained from our numerical
code based on the Dirac gauge in the first row. Below this are the results
obtained from LORENE-rotstar.  Table~\ref{tab:poly} shows that the two sets of results agree
to high accuracy. In particular, the overall discrepancy between the two
different codes is consistent with the errors in the virial identities, which
increase with the rotation frequency. Note that the errors in the virial
identities for the non-rotating configuration listed in the table is about
$10^{-9}$ instead of $10^{-15}$ as shown in Fig.~\ref{fig:polyNR_grv_convg}.
This is due to our choice of using $\epsilon_H=10^{-10}$ in this test.

\begin{table}
\caption{Comparison between our numerical code based on the Dirac gauge (first
  row for each given frequency) and a well-established code LORENE-rotstar
  (second row), which uses a quasi-isotropic gauge for the coordinates, for a
  sequence of $\gamma=2$ polytropic neutron star models  with fixed baryon
  mass $M_b=1.6 M_{\odot}$.  Listed are the rotation frequency $f$,
  gravitational mass $M_g$, total angular momentum parameter $J/M_g^2$, ratio
  of the rotational kinetic energy to the potential energy $T/W$, equatorial
  circumferential radius $R_{eq}$, and  errors indicators in the virial
  identities $GRV2$ and $GRV3$. }\label{tab:poly} 
\begin{indented}
\item[] \begin{tabular}{c c c c c c c}
\hline $f$ (Hz) & $M_g (M_{\odot})$ & $J/M_g^2$ & 
$T/W$ & $R_{eq}$ (km) & $GRV2$ & $GRV3$ \\
\hline
0  & 1.486610961 & 0 & 0 & 14.91222928 & 7e-9 & 2e-9 \\
   & 1.486610965 & 0 & 0 & 14.91222929 & 9e-9 & 1e-9 \\
\hline
100 & 1.486837016 & 0.066036124 & 0.00107222950 & 14.9609161 & 
3e-9 & 3e-8 \\
    & 1.486837013 & 0.066036126 & 0.00107222954 & 14.9609165 & 
3e-8 & 5e-8 \\
\hline
200 & 1.487539907 & 0.13421663 & 0.004395974 & 15.113935 &
2e-7 & 5e-7 \\
    &  1.487539902 &  0.13421665 & 0.004395975 & 15.113937 &
3e-7 & 7e-7 \\
\hline
300 & 1.4888008 & 0.2072617 & 0.010340091 & 15.395905 &
3e-7 & 6e-7 \\
    & 1.4888007 & 0.2072618 & 0.010340095 & 15.395908 &
2e-8 & 6e-7 \\
\hline 
400 & 1.49080352 & 0.289551 & 0.01973704 & 15.865576 &
1e-6 & 3e-6 \\
   & 1.49080359 & 0.289550 & 0.01973701 & 15.865574 & 
7e-7 & 4e-6 \\
\hline 
500 & 1.493991 & 0.390430 & 0.0345952 & 16.67896 & 
4e-6 & 8e-6 \\
  & 1.493990 & 0.390432 & 0.0345954 & 16.67900 &
6e-6 & 1e-5 \\
\hline 
550 & 1.4964095 & 0.455303 & 0.0457202 & 17.35894 & 
2e-5 & 2e-5 \\
   & 1.4964092 & 0.455307 & 0.0457206 & 17.35898  & 
1e-5 & 2e-5 \\
\hline 
600 & 1.500054 & 0.54397 & 0.062368 & 18.5382 &
5e-6 & 2e-6 \\
   & 1.500055 & 0.54398 & 0.062369 & 18.5383 & 
3e-6 & 3e-6 \\
\hline 
$640\approx f_k$ & 1.506928 & 0.695855 & 0.0929183 & 
22.1467 & 2e-5 & 5e-5 \\
    & 1.506929 & 0.695857 & 0.0929188 & 22.1469 &
1e-5 & 6e-5 \\
\hline
\end{tabular}
\end{indented}
\end{table}

\begin{figure}
\centering
\includegraphics*[width=8cm]{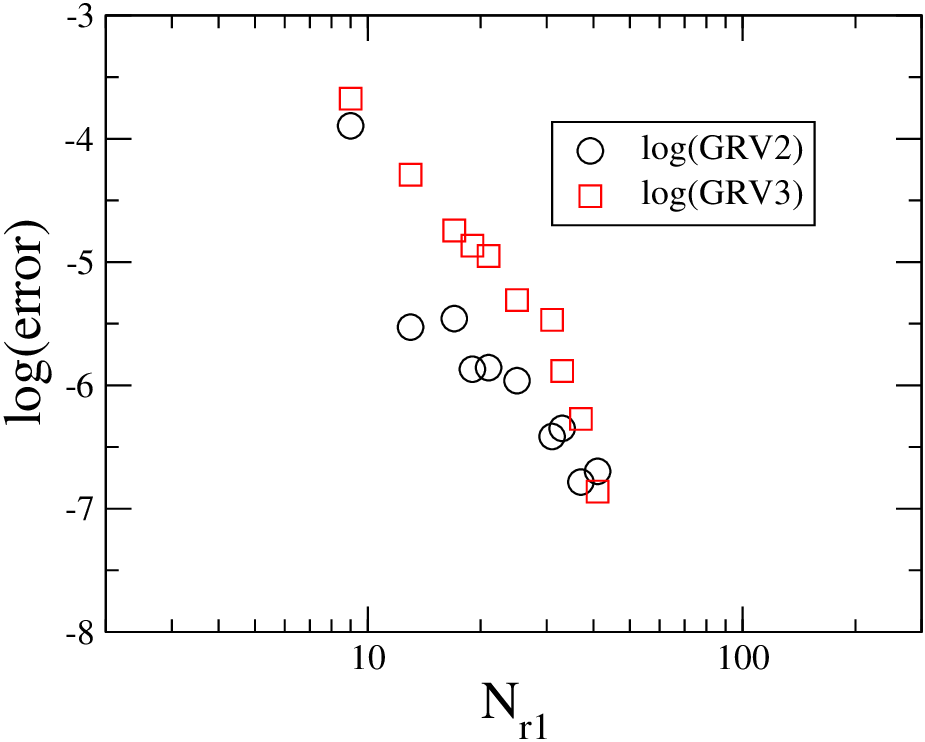}
\caption{Convergence behaviours of $GRV2$ and $GRV3$ for the $f=400$ Hz 
rotating star model listed in Table~\ref{tab:poly}. Note that the plot is in 
log-log scale. The best-fit to the data points shows that 
$GRV2$ ($GRV3$) decreases as $N_{r1}^{-4.3}$ ($N_{r1}^{-3.9}$). }
\label{fig:convg_power}
\end{figure}

\begin{figure}
\centering
\includegraphics*[width=8cm]{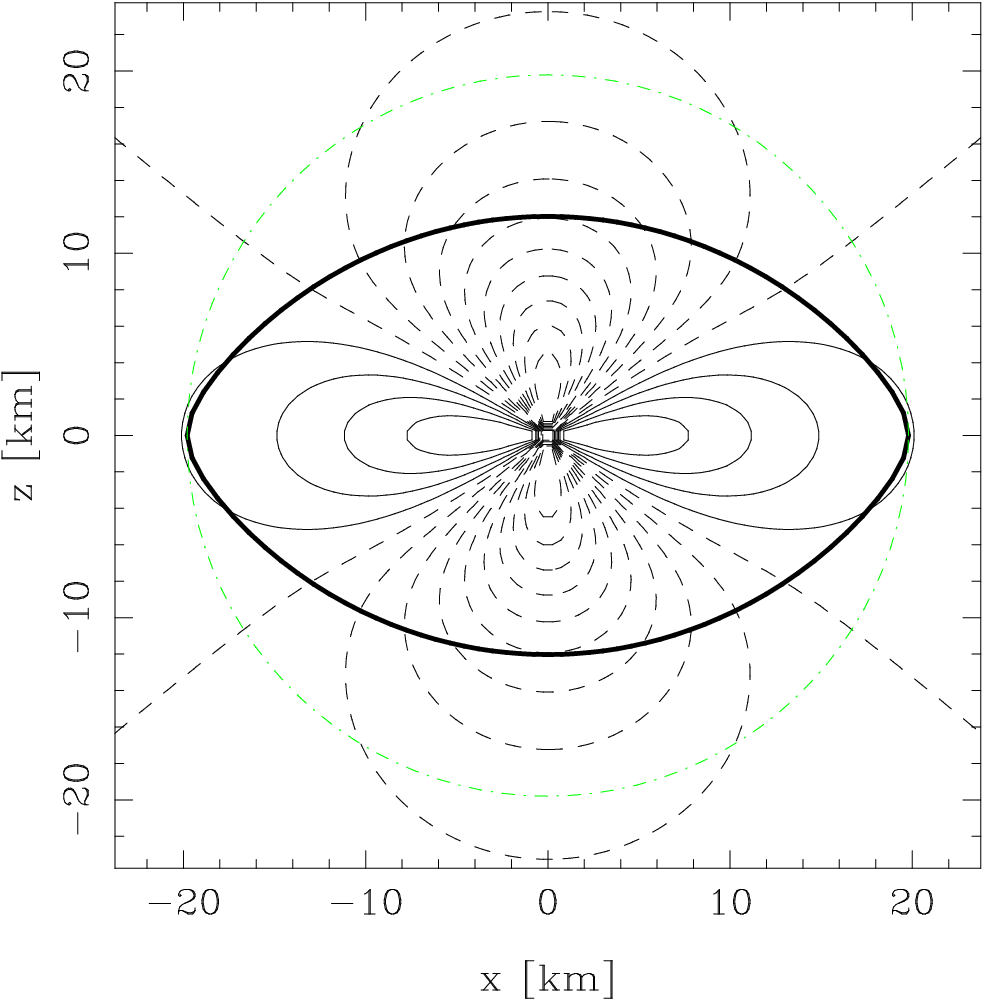}
\caption{Iso-contours of the metric component $h^{\hat r\hat r}$ in the meridional plane
  for the $f=640$ Hz rotating polytropic model given in Table~\ref{tab:poly}.
  The solid (dashed) lines indicate positive (negative) values of the field.
  The thick solid line represents the stellar surface. The dot-dashed circle
  is the boundary between the first two spherical numerical domains.}
\label{fig:poly_rot_hrr}
\end{figure}

\begin{figure}
\centering
\includegraphics*[width=8cm]{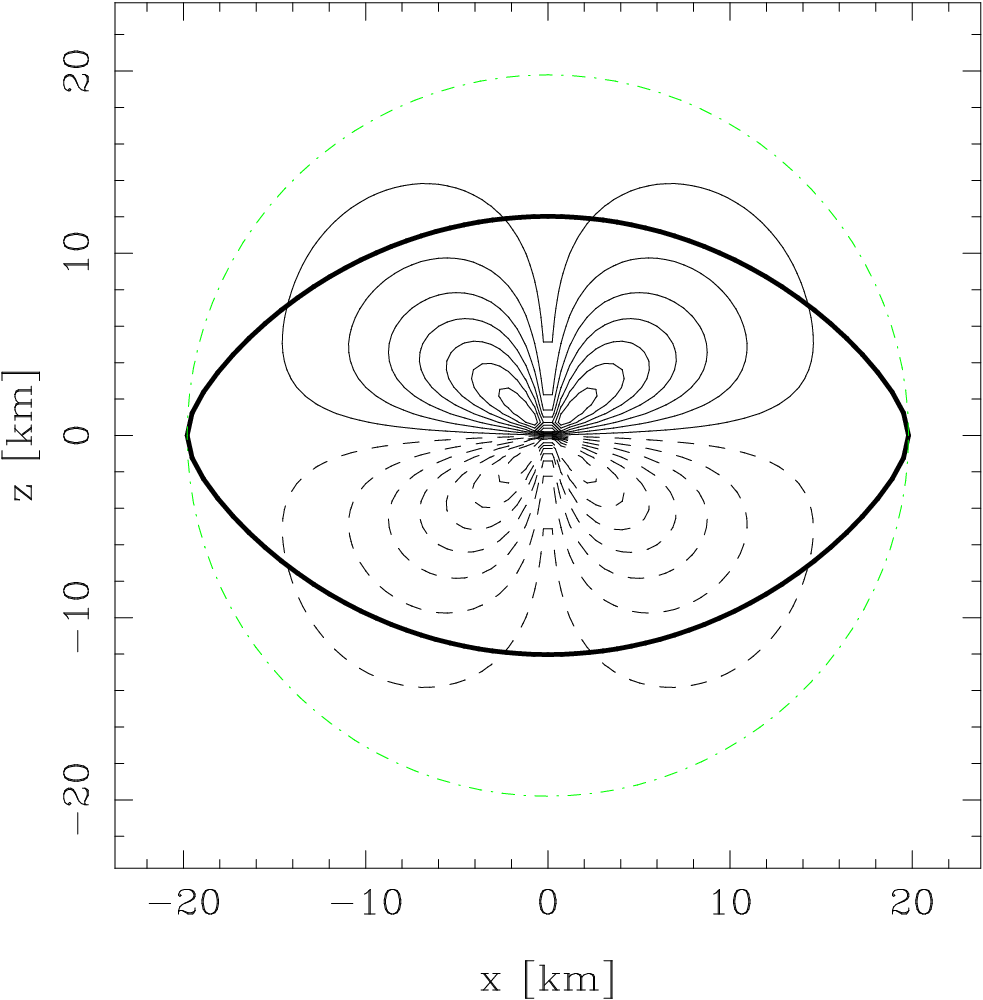}
\caption{Same as Fig.~\ref{fig:poly_rot_hrr} but for the component $h^{\hat r\hat\theta}$.}
\label{fig:poly_rot_hrt}
\end{figure}

\begin{figure}
\centering
\includegraphics*[width=8cm]{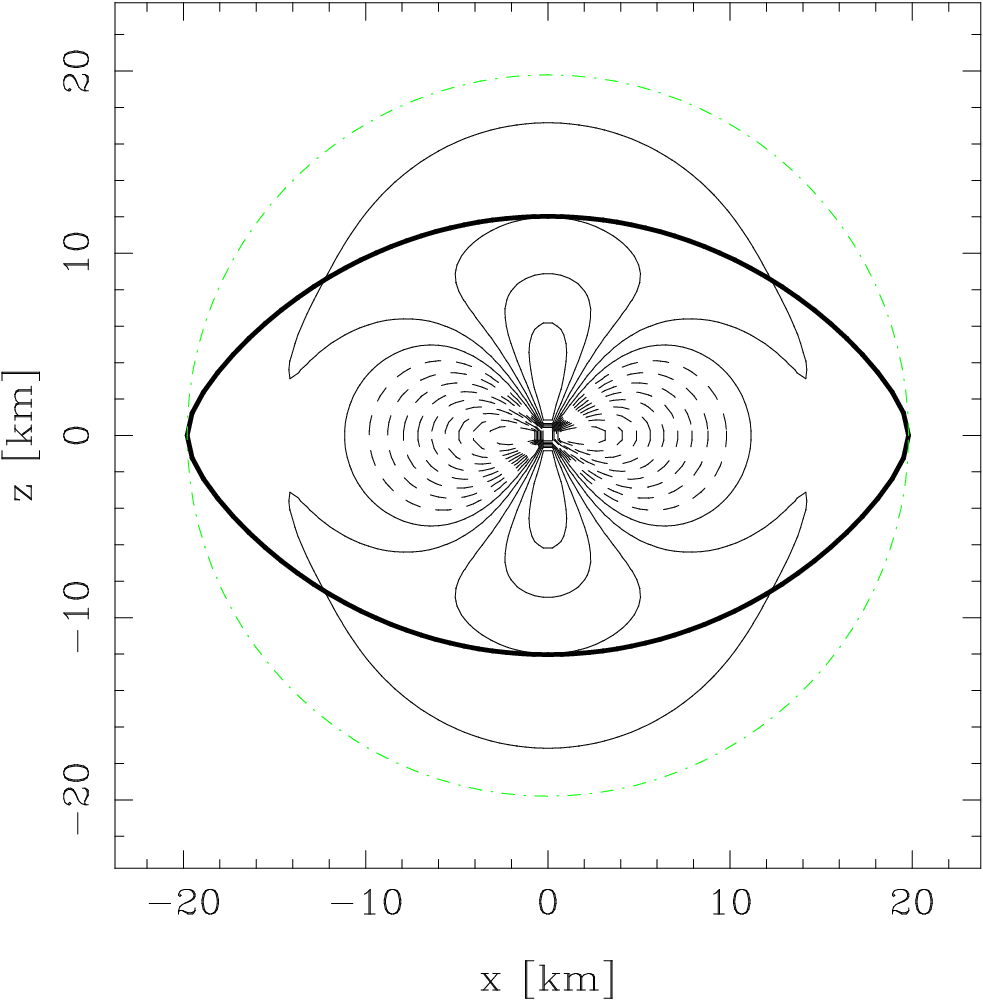}
\caption{Same as Fig.~\ref{fig:poly_rot_hrr} but for the component 
$h^{\hat\theta\hat\theta}$. }
\label{fig:poly_rot_htt}
\end{figure}

\begin{figure}
\centering
\includegraphics*[width=8cm]{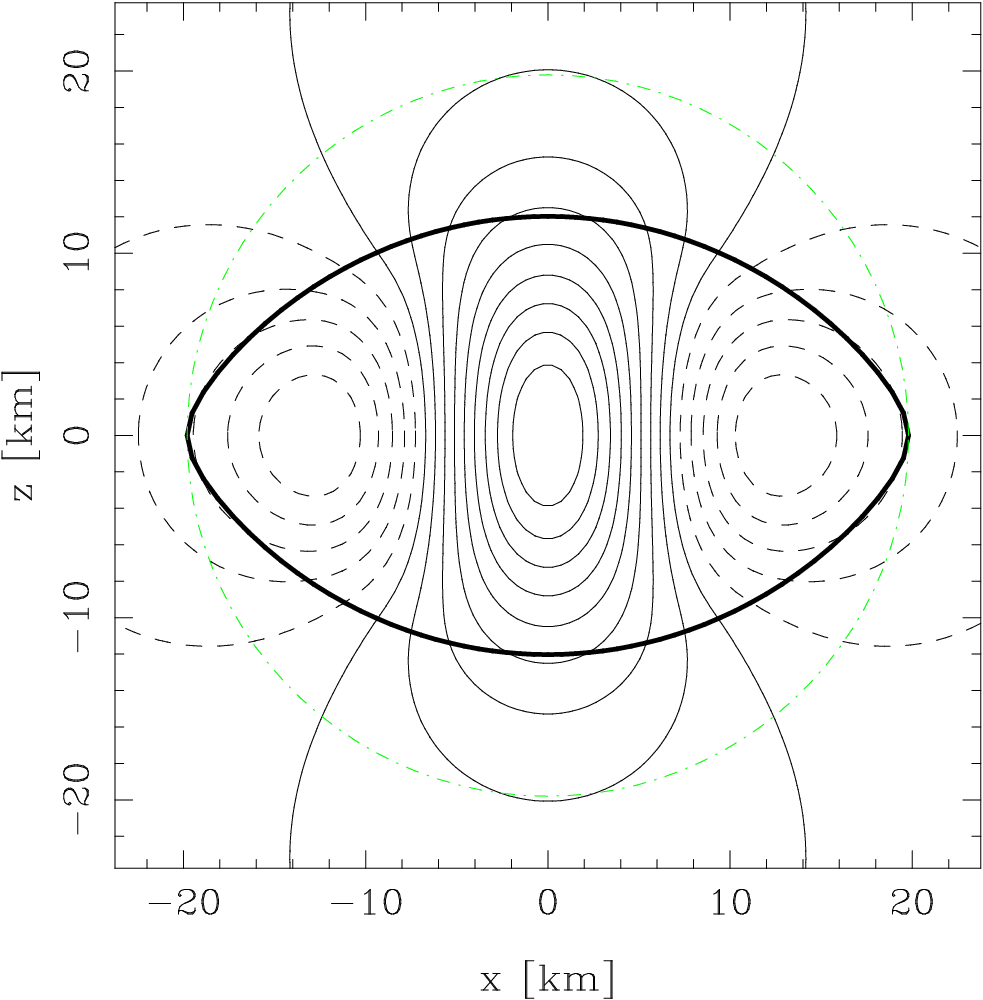}
\caption{Same as Fig.~\ref{fig:poly_rot_hpp} but for the component 
$h^{\hat\varphi\hat\varphi}$. }
\label{fig:poly_rot_hpp}
\end{figure}

We note, however, that the numerical error no longer decreases exponentially
with the number of grid points as in the non-rotating case, but as a 
power-law (see Fig.~\ref{fig:convg_power}), due to the
discontinuities in the derivative of the matter fields at the stellar surface.
The non-rotating model is free from any such phenomenon because the stellar
surface is at the boundary between two spherical numerical domains.  For
rotating models, because of the flattening of the stars, the stellar surface
no longer coincides with the boundary of the domains, and hence the spectral
method loses its exponential-convergence property.  Such difficulty associated
with rotating stars can be handled by the adaptation of the numerical domains
to the stellar surface as developed in \cite{bon98}. We plan to improve our
numerical code by implementing this surface-adaptation technique in the near
future\footnote{Such numerical technique is already available in
  LORENE-rotstar as described in \cite{gou99}. However, in order to compare to
  our numerical code, the results obtained from LORENE-rotstar as listed in
  Table~\ref{tab:poly} are based on fixed spherical numerical domains.}.
Nevertheless, even without a surface-adaptation technique,
Table~\ref{tab:poly} shows that both numerical codes still agree to high
accuracy and achieve a precision of $10^{-5}$ for a configuration rotating
near the mass-shedding limit (i.e., the $f=640\approx f_k$ Hz configuration in
Table~\ref{tab:poly}, where $f_k$ is the Kepler frequency).  
To visualise the
gravitational field generated by a rotating star, we plot in
Figs.~\ref{fig:poly_rot_hrr}-\ref{fig:poly_rot_hpp} the non-vanishing
components for the metric field $h^{ij}$ (namely, $h^{\hat r\hat r}$, $h^{\hat
  r\hat \theta}$, $h^{\hat \theta \hat\theta}$, and $h^{\hat\varphi
  \hat\varphi}$) for the rotating star model with $f=640$ Hz.  In these
figures, we show the iso-contours of the fields in the meridional plane, where
solid (dashed) lines indicate positive (negative) values of the fields. The
thick solid lines represent the stellar surface. Finally, the dot-dashed
circles represent the boundary between the first two spherical numerical
domains.  The figures show clearly that the gravitational field is dominated
by the quadrupole moment.

\begin{figure}
\centering
\includegraphics*[width=8cm]{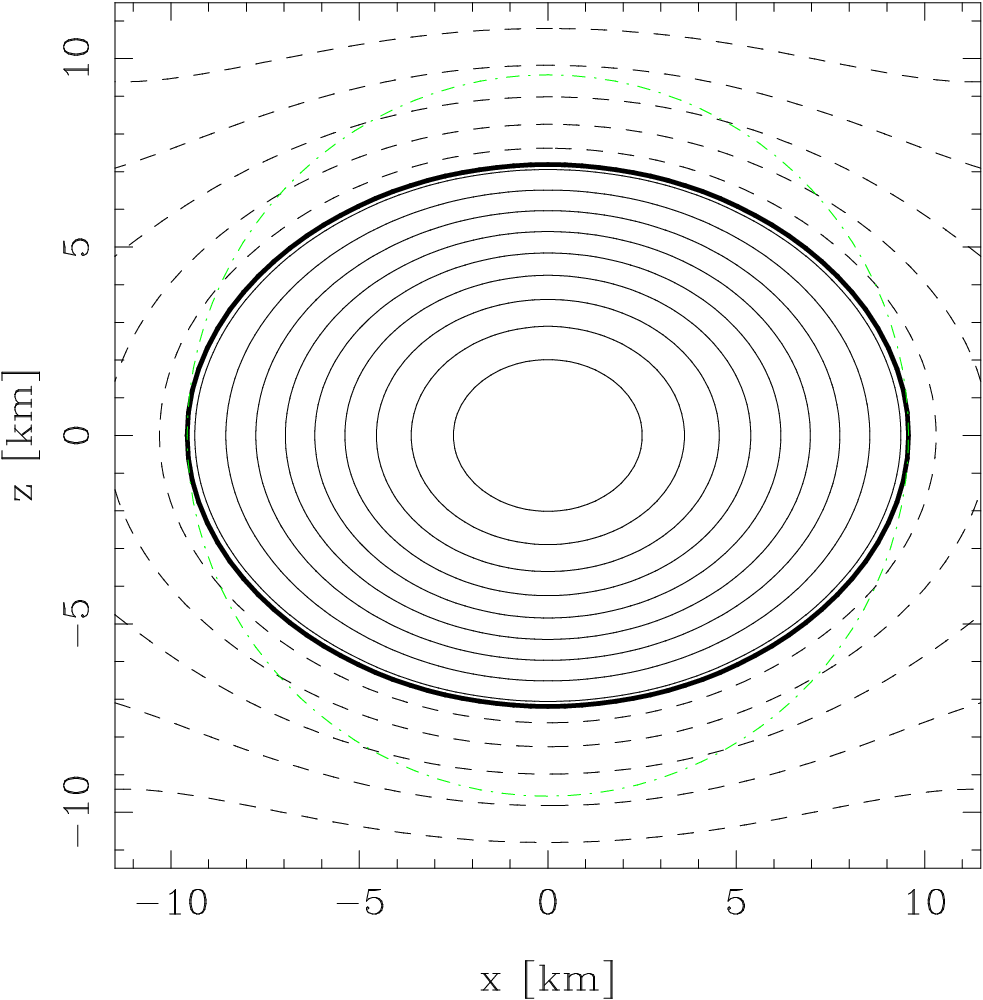}
\caption{Meridional plane cross section of a rapidly rotating strange star. The star has 
a baryon mass $M_b=2.3 M_{\odot}$, gravitational mass $M_g=1.787M_{\odot}$, and rotation
frequency $f=1000$ Hz. The lines are iso-contours of the log-enthalpy $H$. The thick solid
line represents the stellar surface. Outside the star, $H$ is defined by the first integral 
of motion Eq.~(\ref{eq:fluid_1st}).}
\label{fig:mit_rot_H}
\end{figure}

To further calibrate our numerical code against LORENE-rotstar, we now compare
the two codes using a very relativistic and rapidly rotating strange star
model. The matter is described by the MIT bag model as described in
Sec.~\ref{sec:eos}. This configuration, shown in Fig.~\ref{fig:mit_rot_H}, has
a baryon mass $M_b=2.2 M_{\odot}$ and compactness ratio $M_g/R_{eq}=0.204$
(with the gravitational mass $M_g=1.719M_{\odot}$ and the circumferential
equatorial radius $R_{eq}=12.425$ km). The rotation frequency is $f=1000$ Hz.
In Table~\ref{tab:str}, we show the values of various physical quantities obtained from
both numerical codes, together with the relative difference between them. As
in the case of the polytropic EOS model, the discrepancy of the two numerical
codes is consistent with the errors in $GRV2$ and $GRV3$.  Note also that,
even with a strong density discontinuity at the strange star surface, our
numerical model still achieves a precision of $10^{-3}$.

\begin{table}
\caption{Comparison between our numerical code (LORENE-rotstar\_dirac) and LORENE-rotstar for a
rapidly rotating strange star model. The star has a baryon mass $M_b=2.2M_{\odot}$, gravitational 
mass $M_g=1.719M_{\odot}$, and rotation frequency $f=1000$ Hz. }\label{tab:str}
\begin{indented}
\item[] \begin{tabular}{c c c c}
\hline    &  rotstar\_dirac  & rotstar  & rel. diff. \\
\hline
$M_g (M_{\odot})$ & 1.7194 & 1.7198  & 2e-4 \\
$J/M_g^2$ & 0.5940 & 0.5945 & 8e-4 \\
$T/W$  & 0.0888 & 0.0890 &  2e-3 \\
$R_{eq}$ (km) & 12.425 & 12.433 & 7e-4 \\ 
$GRV2$ & 7e-4 & 2e-4  &  \\
$GRV3$ & 1e-3 & 6e-4 &  \\
\hline
\end{tabular}
\end{indented}
\end{table}

\section{Conclusion}

In this paper we have developed a computer code (LORENE-rotstar\_dirac) to
construct relativistic rotating stars within the framework of a new
constrained-evolution formulation of the $3+1$ Einstein equations based on the
Dirac gauge and maximal slicing \cite{bon04}.  As the Dirac gauge fixes the
spatial coordinates on each time slices, including the initial one, this
formulation can be used to compute stationary solutions of the Einstein
equations simply by setting the time derivative terms in various equations to
zeros.  The system reduces to two scalar elliptic equations for the lapse
function $N$ and conformal factor $\Psi$ (equivalently for $Q:= \Psi^2 N$), a
vectorial elliptic equation for the shift vector $\beta^i$, and a tensorial
elliptic equation for a tensor field $h^{ij}$.  We couple this system of
equations to the first integral of motion for the matter, and solve the
equations iteratively using multi-domain spectral method.

We have demonstrated that this formulation can be used to compute stationary
rotating equilibrium configurations to high accuracy. In particular, we used
the polytropic EOS and MIT bag model to calculate rotating neutron star and
strange star models respectively.  We compared our code to a well-established
code LORENE-rotstar, which uses a quasi-isotropic gauge to fix the
coordinates, and found that the global quantities of the numerical models
obtained from the two codes agree to high accuracy. The discrepancy between
the two codes is consistent to the errors in the virial identities.

Finally, we remark that the proposed constrained-evolution scheme \cite{bon04}
is particular well suited to the conformally-flat relativistic hydrodynamics
code, with a metric solver based on spectral methods and spherical
coordinates, developed by Dimmelmeier {\it et al.} in the so-called {\it
  Mariage des Maillages} (MDM) project \cite{dim05}. The numerical code that
we described in this paper can be used to generate rotating-star initial data
for hydrodynamics simulations in full general relativity within the new
constrained-evolution scheme \cite{bon04} for the MDM project.

\ack We thank Silvano Bonazzola, \'Eric Gourgoulhon and Philippe
Grandcl\'ement for helpful discussions and hints for this work. L.M.L is
supported by a Croucher Foundation fellowship.

\appendix
\section{Resolution of the Poisson equations for $h^{ij}$}
\label{sec:appen}

Here we describe the numerical strategy used to solve the tensorial Poisson
equation ~(\ref{eq:hij}), imposing that the solution $h^{ij}$ satisfies the 
gauge condition~(\ref{eq:dirach}) and be such that the conformal metric has a
unitary determinant:
\begin{equation}
  \label{eq:detgam1}
  \det \left( \tilde{\gamma}^{\hat i\hat j} = f^{\hat i\hat j} + h^{\hat i\hat j} \right) = 1 .
\end{equation}
Note that this relation follows directly from the definition of the conformal
factor in the proposed constrained scheme (see Eqs.~(\ref{eq:conf_gam}) and
(\ref{eq:conf_psi})), together with the condition $\det f_{\hat i\hat j} = 1$
in the orthonormal basis $(\w{e}_{\hat i})$ (see Sec.~\ref{sec:global}).  In
Ref.~\cite{bon04}, one would solve two (scalar) Poisson equations: for
$h^{\hat r\hat r}$ and the potential $\mu$ (see Eq.~(\ref{eq:defmu})); the other four
components are deduced from the three gauge conditions and the non-linear
relation (\ref{eq:detgam1}) through an iteration. The drawback of this method
is that some components of $h^{ij}$ are calculated as second radial
derivatives of $h^{\hat r\hat r}$ and $\mu$. Since the source of Eq.~(\ref{eq:hij})
contains second-order radial derivatives of $h^{ij}$, one needs to calculate
fourth-order radial derivatives of $h^{\hat r\hat r}$ and $\mu$, which are solutions of
scalar-like Poisson equations with matter terms on the RHS.  In the case of
neutron stars, it is quite often that radial density profiles have a
discontinuous derivative at the surface of the star.  Therefore, $h^{\hat r\hat r}$ and
$\mu$ admit discontinuous third-order radial derivatives and their
fourth-order derivatives cannot be represented at all by means of spectral
methods. A solution could be to use adaptive mapping: the boundary between
two spectral domains coincides with the (non-spherical) surface of the star
(see \cite{bon98}). Still, the evaluation of a fourth-order radial derivative
introduces too much numerical noise, even using spectral methods. We have
therefore chosen to use a different approach, detailed hereafter.

Instead of using directly all the spherical components of the tensor $h^{ij}$,
we use only the $\hat r\hat r$- component, the trace $h = f_{ij} h^{ij}$ and the four potentials $\eta$,
$\mu$, $W$ and $X$ defined as follow, in the orthonormal basis $( \w{e}_{\hat
  i})$:
\begin{eqnarray}
  \label{eq:defetamu}
    h^{\hat r\hat \theta} & = & {1\over r} \left( \frac{\partial\eta}{\partial\theta} - {1\over\sin\theta}
        \frac{\partial\mu}{\partial\varphi} \right), \\
      h^{\hat r\hat \varphi} & = & {1\over r} \left( {1\over\sin\theta} \frac{\partial\eta}{\partial\varphi}
  + \frac{\partial\mu}{\partial\theta} \right),
\end{eqnarray}
and 
\begin{eqnarray}
  \label{eq:defWX}
  P &=& \frac{\partial^2 W}{\partial\theta^2} - \frac{1}{\tan \theta}
  \frac{\partial W}{\partial\theta} - \frac{1}{\sin^2 \theta} \frac{\partial^2
  W}{\partial\varphi^2} - 2 \frac{\partial}{\partial\theta} \left(
  \frac{1}{\sin \theta} \frac{\partial X}{\partial\varphi} \right) ,\\
  h^{\hat \theta\hat \varphi} &=& \frac{\partial^2 X}{\partial\theta^2} - \frac{1}{\tan
  \theta} \frac{\partial X}{\partial\theta} - \frac{1}{\sin^2 \theta}
  \frac{\partial^2 X}{\partial\varphi} + 2 \frac{\partial}{\partial\theta}
  \left( \frac{1}{\sin \theta} \frac{\partial W}{\partial\varphi} \right);
\end{eqnarray}
with $P= ( h^{\hat \theta\hat \theta} - h^{\hat \varphi\hat \varphi} ) / 2$. These equations can
be inverted to compute the potentials in terms of the angular part of the
Laplace operator 
\begin{equation}
  \label{eq:lapang}
  \Delta_{\theta\varphi} = \frac{\partial^2}{\partial \theta^2} +
  \frac{1}{\tan \theta} \frac{\partial}{\partial \theta} + \frac{1}{\sin^2
  \theta} \frac{\partial^2}{\partial \varphi^2} ,
\end{equation}
giving
\begin{eqnarray}
\Delta_{\theta \varphi} \eta& =&r \left( \frac{\partial h^{\hat r\hat \theta}}{\partial
  \theta} + \frac{h^{\hat r\hat \theta}}{\tan \theta} + \frac{1}{\sin\theta}
  \frac{\partial h^{\hat r\hat \varphi}}{\partial \varphi} \right),\\
\Delta_{\theta \varphi} \mu & = &r \left( \frac{\partial h^{\hat r\hat \varphi}}{\partial
  \theta} + \frac{h^{\hat r\hat \varphi}}{\tan \theta} - \frac{1}{\sin\theta}
  \frac{\partial h^{\hat r\hat \theta}}{\partial \varphi} \right), \label{eq:defmu}\\
\Delta_{\theta\varphi} \left( \Delta_{\theta\varphi} + 2 \right) W & = &
  \frac{\partial^2 P}{\partial \theta^2} + \frac{3}{\tan \theta} \frac{\partial
  P}{\partial \theta} - \frac{1}{\sin^2 \theta} \frac{\partial^2 P}{\partial \varphi^2} -2P
  \nonumber \\ && +
  \frac{2}{\sin \theta} \frac{\partial }{\partial \varphi} \left( \frac{\partial
  h^{\hat \theta\hat \varphi}}{\partial \theta} + \frac{h^{\hat \theta\hat \varphi}}{\tan \theta} \right),\\ 
\Delta_{\theta\varphi} \left( \Delta_{\theta\varphi} + 2 \right) X &=&
  \frac{\partial^2 h^{\hat \theta\hat \varphi}}{\partial \theta^2} + \frac{3}{\tan \theta}
  \frac{\partial h^{\hat \theta\hat \varphi}}{\partial \theta} - \frac{1}{\sin^2 \theta}
  \frac{\partial^2 h^{\hat \theta\hat \varphi}}{\partial \varphi^2} -2h^{\hat \theta\hat \varphi}
  \nonumber\\ && -
  \frac{2}{\sin \theta} \frac{\partial }{\partial \varphi} \left( \frac{\partial
  P}{\partial \theta} + \frac{P}{\tan \theta} \right). 
\end{eqnarray}

These quantities $\eta, \mu, W, X$ are interesting for, at least, two reasons:
first they can be expanded onto a basis of scalar spherical harmonics
$Y_\ell^m(\theta, \varphi)$, which are often used in the framework of spectral
methods, for they are eigenfunctions of the angular Laplace operator
(\ref{eq:lapang}). Furthermore, the three gauge conditions (\ref{eq:dirach})
can be reformulated in terms of these potentials:
\begin{eqnarray}
  \frac{\partial h^{\hat r\hat r}}{\partial r} + \frac{3h^{\hat r\hat r}}{r} + \frac{1}{r^2}
  \Delta_{\theta\varphi} \eta - \frac{h}{r} = 0,\label{e:dir1}\\ 
\frac{\partial \eta}{\partial r} + \frac{2\eta}{r} + \left( \Delta_{\theta\varphi} + 2
\right) W + \frac{1}{2} \left( h - h^{\hat r\hat r} \right) = 0 ,\label{e:dir2}\\
\frac{\partial \mu}{\partial r} + \frac{2\mu}{r}+ \left( \Delta_{\theta\varphi} + 2
\right) X = 0 .\label{e:dir3}
\end{eqnarray}
When decomposing the fields on a basis of spherical harmonics, these relations
reduce to a system of ordinary differential equations with respect to $r$,
which is solved by spectral methods in a similar way to the Poisson equation
\cite{gra01}.

The numerical algorithm is then:
\begin{enumerate}
\item transform the source term of Eq.~(\ref{eq:hij}) to the Cartesian basis,
\item solve the resulting six decoupled scalar Poisson equations for $h^{ij}$,
\item transform $h^{ij}$ back to the spherical basis and compute the
  potentials $W$ and $X$,
\item do an iteration on $h$, first solving the
  system~(\ref{e:dir1})-(\ref{e:dir3}) with $h, W$ and $X$ as sources and then
  calculating the new value of $h$ from the non-linear equation~(\ref{eq:detgam1}).
\end{enumerate}
Since the system is overdetermined (four additional relations to satisfy), the
integrability condition is that the source of the tensor Poisson
equation~(\ref{eq:hij}) be divergence-free. We do not impose this condition
during the main iteration of the code, since this is not true for intermediate
solutions of the metric and matter fields. We only check that this is
satisfied, up to the accuracy of the code, at the end of the iteration, and we
have found that this was true at the error level given by the virial
identities (see Sec.~\ref{sec:global}). The potentials $W$ and $X$ have been
chosen among the six degrees of freedom of $h^{ij}$ because none of their
radial derivatives appear in the gauge
conditions~(\ref{e:dir1})-(\ref{e:dir3}), hence, we do not calculate any
radial derivative of these quantities to get the other components of $h^{ij}$.
Another reason for this choice is that, when considering a more general case
of dynamically evolving spacetime, these two potentials are asymptotically
related to the two gravitational wave polarisation modes: $P \to h_+$ and
$h^{\theta\varphi} \to h_\times$ in our asymptotically transverse-traceless
gauge\footnote{The condition (\ref{eq:detgam1}) implies that $h=0$ to the
  linear order}. Note that in our case of stationary and axisymmetric
spacetime, we have $\mu = X = 0$, which simplifies the resolution.

\bibliographystyle{prsty}

\end{document}